\documentclass[prl,twocolumn,amsmath,amssymb,10pt,aps,longbibliography,superscriptaddress,citeautoscript,bibnotes,footinbib]{revtex4-2}

\usepackage{feynmp}
\usepackage{amsmath}
\usepackage{graphicx}
\usepackage{multirow}
\usepackage{latexsym}
\usepackage{textcomp}
\usepackage{verbatim}
\usepackage{color}
\usepackage{bm}
\usepackage{subfigure}
\usepackage{everysel}
\usepackage{keyval}
\usepackage{ragged2e}
\usepackage{dsfont}
\usepackage{amssymb}
\usepackage{enumitem}
\usepackage{pstricks}
\usepackage{mathtools}

\usepackage{braket}
\usepackage{slashed} 
\usepackage{hyperref}        
\hypersetup{
     colorlinks=true,
     citecolor = red,    
     linkcolor=magenta,
     }

\usepackage{graphicx}
\usepackage{xcolor}
\usepackage{amsmath}
\usepackage{bbold}
\usepackage{amsfonts}
\usepackage{bbm}
\usepackage{comment}

\newcommand{\osum}{{%
    \setbox0\hbox{\circ}%
    \rlap{\hbox to \wd0{\hss\sum\hss}}\box0
}}

\begin{document}

\title{Cavity magnonics with easy-axis ferromagnet: critically enhanced magnon squeezing and light-matter interaction}

\author{Jongjun M. Lee}
\affiliation{Department of Physics, Pohang University of Science and Technology (POSTECH), Pohang 37673, Korea}

\author{Hyun-Woo Lee}
\thanks{Electronic Address: hwl@postech.ac.kr}
\affiliation{Department of Physics, Pohang University of Science and Technology (POSTECH), Pohang 37673, Korea}

\author{Myung-Joong Hwang}
\thanks{Electronic Address: myungjoong.hwang@duke.edu}
\affiliation{Division of Natural and Applied Sciences, Duke Kunshan University, Kunshan, Jiangsu 215300, China}
\affiliation{Zu Chongzhi Center for Mathematics and Computational Science, Duke Kunshan University, Kunshan, Jiangsu 215300, China}

\begin{abstract}
Generating and probing the magnon squeezing is an important challenge in the field of quantum magnonics. In this work, we propose a cavity magnonics setup with an easy-axis ferromagnet to address this challenge. To this end, we first establish a mechanism for the generation of magnon squeezing in the easy-axis ferromagnet and show that the magnon squeezing can be critically enhanced by tuning an external magnetic field near the Ising phase transition point. When the magnet is coupled to the cavity field, the effective cavity-magnon interaction becomes proportional to the magnon squeezing, allowing one to enhance the cavity-magnon coupling strength using a static field. We demonstrate that the magnon squeezing can be probed by measuring the frequency shift of the cavity field. Moreover, a magnonic superradiant phase transition can be observed in our setup by tuning the static magnetic field, overcoming the challenge that the magnetic interaction between the cavity and the magnet is typically too weak to drive the superradiant transition. Our work paves the way to develop unique capabilities of cavity magnonics that goes beyond the conventional cavity QED physics by harnessing the intrinsic property of a magnet.
\end{abstract}
 
\date{\today}
\maketitle


{\it Introduction.---} 
The magnon is a quantized spin wave excitation of the spin ordering in magnets~\cite{kittel1996introduction,blundell2003magnetism,chumak2012direct,cornelissen2015long,xing2019magnon,li2020spin} and has many virtues. Its long lifetime and negligible Joule heating make the magnon a potential candidate for spin-based computation and communication, which has led to the development of a new research field called magnonics~\cite{chumak2015magnon,chumak2022roadmap}. Various magnonic device components have been developed~\cite{chumak2017magnonic,wang2020magnonic,wang2020nonlinear,haldar2021functional}. The magnon can also have a non-trivial topological order and exhibit anomalous transport phenomena such as the magnon Hall effect and the magnon Nernst effect~\cite{zhang2013topological,chisnell2015topological,owerre2016first,nakata2017magnonic,chen2018topological,yao2018topological,kim2022topological}. 
However, all these properties can be understood without quantizing the spin wave.

Recent studies unveiled various exotic properties of the magnon arising from its bosonic quantization. In particular, squeezing of a quantized magnon state~\cite{kamra2020magnon} gives rise to various exotic properties such as the spin noise reduction below the standard quantum limit~\cite{zhao2004magnon,zhao2006magnon}, the magnon spin angular momentum enhancement beyond the standard value $\hbar$~\cite{kamra2016super}, the entanglement generation~\cite{kamra2019antiferromagnetic,zou2020tuning,mousolou2021magnon}, and the macroscopic quantum state generation (cat state)~\cite{sharma2021spin,sun2021remote,kounalakis2022analog}. Other interesting phenomena involving magnons such as anti-bunching~\cite{xie2020quantum,yuan2020magnon,li2021tunable,li2022tunable}, the Bose-Einstein condensation~\cite{nikuni2000bose,demokritov2006bose,giamarchi2008bose,bunkov2010magnon,zapf2014bose,serga2014bose}, the Casimir effect~\cite{neuberger1989finite,cheng2018interlayer,nakata2023magnonic}, and superradiant phase transitions (SPTs)~\cite{roman2021photon,bamba2022magnonic,mckenzie2022theory,liu2023switchable} are also reported.
Unfortunately, very few of these exotic phenomena are experimentally verified~\cite{zhao2004magnon,zhao2006magnon,demokritov2006bose}. 
Considering that they can be useful resources for quantum computation and communication, it is highly desired to bring those exotic theoretical possibilities into the realm of experimentally testable regimes.

In this Letter, we propose a cavity magnonics setup with an easy-axis ferromagnet as an experimentally promising platform to realize and probe the magnon squeezing and associated effects. The key feature of our setup is that simply applying an external \emph{static} magnetic field to the magnet can critically enhance both magnon squeezing and cavity-magnon interaction. This enhancement creates opportunities in two directions. Firstly, since the magnon squeezing characteristics are imprinted on the cavity photons through the enhanced cavity-magnon interaction, the magnon squeezing and the enhanced spin angular momentum can be probed through the cavity photon measurement, which can be readily performed with quantum optical toolboxes available for cavity and circuit quantum electrodynamics (QED) systems~\cite{schuster2007resolving,Hofheinz2009a,Lolli2015a,Blais2020a}. This probing scheme is expected to be more robust to the effects of noises than transport-based probing schemes of the magnon squeezing~\cite{kamra2016super,kamra2019antiferromagnetic}. Secondly, it allows one to induce an SPT by tuning the static field even though a bare cavity-magnon interaction is weak. 
Since our magnon-squeezing-based scheme to effectively enhance the interaction does not enhance the $A^{2}$ term, the term remains small, and does not significantly contribute to nor prevent the SPT. Moreover, our scheme is drastically different from existing schemes where the enhanced interaction is realized in a rotating frame~\cite{leroux2018enhancing,Qin2018a,Zhu2020a} using time-dependent parametric driving. Our scheme utilizes the intrinsic property of the magnet without the parametric driving; therefore, it is free of unwanted processes due to the driving field that may limit the experimentally achievable coupling strength~\cite{Braumueler2017a,Langford2017a}, demonstrating unique capabilities of the cavity magnonics.


In the absence of the cavity, we first establish that the interplay between the static magnetic field and the magnetic anisotropy leads to the magnon squeezing of an easy-axis ferromagnet. As the field is tuned to drive the magnet close to the Ising transition, both the squeezing and the magnon spin angular momentum are critically enhanced with a power law. When the magnet is coupled to the magnetic field of the cavity photons in addition to the classical external field, the diverging spin angular momentum induces the diverging fluctuation of the cavity field, which in turn results in a ground-state superradiance where the $Z_2$ symmetry associated with the mirror reflection symmetry is spontaneously broken. We study the critical behavior of the energy spectrum and the squeezing for the cavity magnon-polariton and identify its universality class to be that of fully connected systems.

\begin{figure}
    \centering
    \includegraphics[width=\linewidth]{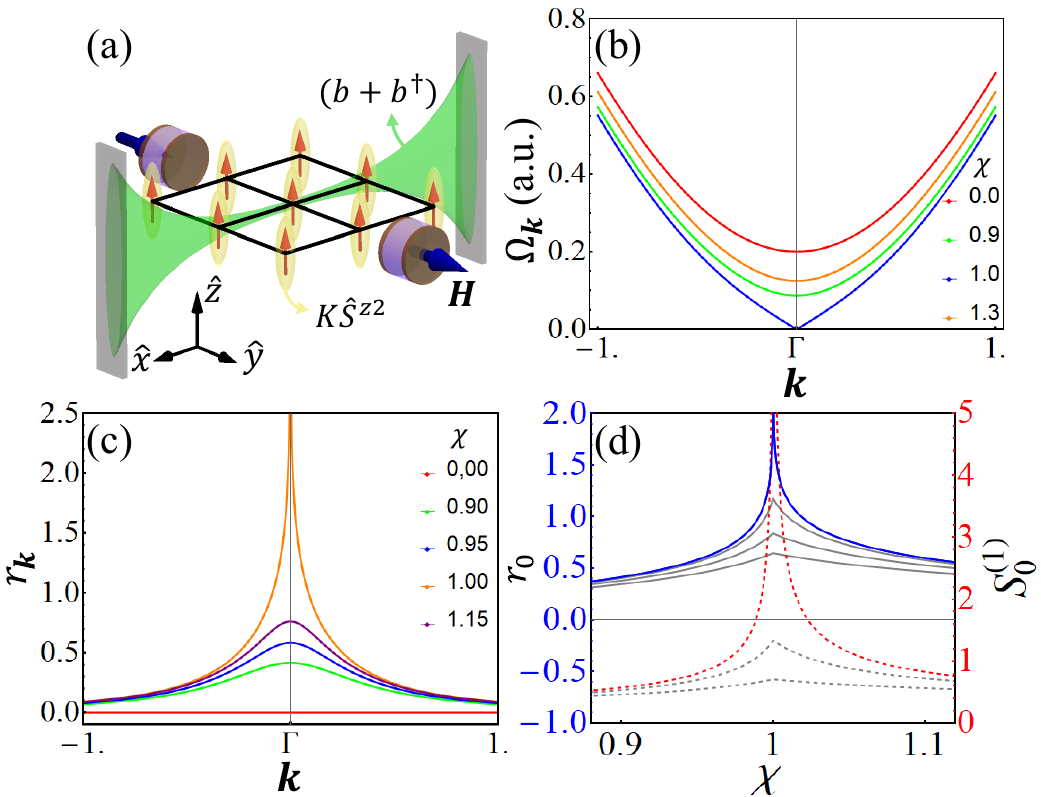}
    \caption{(a) Schematic illustration of our cavity-magnet system. The green region enclosed by the rectangles, the arrows on the grid, and the cylinders represent the cavity field, the easy-axis ferromagnet, and the coil for the external field, respectively. (b) Magnon dispersions along the momentum $\bm{k}$ for a few values of $\chi$ near the $\Gamma$ point ($\bm{k}=0$). 
    (c) Squeezing parameters along the momentum $\bm{k}$ for a few values of $\chi$ near the $\Gamma$ point. 
    (d) Squeezing parameters and spin angular momentums of the magnon along $\chi$ near the critical point. The gray lines represent the quantities with finite momentum.
    }
    \label{FIG_1}
\end{figure}

{\it Model Hamiltonian.---} 
The cavity magnonics setup [Fig.~\ref{FIG_1}(a)] consists of an easy-axis ferromagnetic insulator subject to an external static magnetic field, coupled to a cavity magnetic field. We model its low-energy excitations with the Hamiltonian, 
\begin{equation}
\begin{aligned}
    \mathcal{H}_{\text{tot}} &= \omega b^{\dagger}b - \frac{g}{\sqrt{N}}\sqrt{\frac{2}{S}} (b+b^{\dagger})\sum_{\bm{r}}  \hat{S}^{z}_{\bm{r}}+ \mathcal{H}_{\text{FM}},
\end{aligned}
\label{Eq_H_tot}
\end{equation}
with the spin Hamiltonian part of the ferromagnet,
\begin{equation}
    \mathcal{H}_{\text{FM}} = -J \sum_{\langle \bm{r}, \bm{r'}\rangle} \hat{\bm{S}}_{\bm{r}} \cdot \hat{\bm{S}}_{\bm{r'}}
     - \sum_{\bm{r}}(K \hat{S}^{z 2}_{\bm{r}} + \gamma \mu \bm{H} \cdot \hat{\bm{S}}_{\bm{r}} ),
\label{Eq_H_FM}
\end{equation}
where $g=\gamma \mu h\sqrt{NS/2}$ is the coupling strength, $N$ is the number of spins, $\hat{\bm{S}}_{\bm{r}}$ denotes the dimensionless spin of magnitude $S$ at site $\bm{r}$, $b$ is a bosonic operator of the cavity field with the frequency $\omega$ and the strength $h$, $\mu$ is the permeability, and $\bm{H}$ is the static field. $J>0$ denotes the Heisenberg-type isotropic ferromagnetic exchange interaction between the neighboring sites, $K>0$ represents the easy-axis anisotropy along the $z$ axis, and $\gamma$ is the gyromagnetic ratio for the Zeeman interaction. $J$ has the dominant energy scale responsible for the ferromagnetic spin ordering, whose direction is determined by the competition between $K$, $\bm{H}$, and $g$ with weak characteristic energy scales. This competition gives rise to a phase transition with the diverging magnonic squeezing and the ground-state superradiance. For simplicity, we consider $\bm H=H_0 \hat y$ throughout the paper. The total Hamiltonian then has a $Z_2$ symmetry corresponding to the following transformation
\begin{align}
     (b, \hat S_r^x,\hat S_r^y,\hat S_r^z)\xrightarrow{\mathcal{M}_y}(-b, -\hat S_r^x, \hat S_r^y, -\hat S_r^z),
 \end{align}
where $\mathcal{M}_{y}$ is the mirror reflection operator along the $y$ axis. As we show below, upon modulating the static magnetic field $H_0$, the spontaneous breaking of $\mathcal{M}_y$ occurs, which leads to the superradiant phase.

{\it Critically enhanced magnon squeezing.---} 
We first investigate the spin Hamiltonian $\mathcal{H}_{\text{FM}}$ [Eq.~(\ref{Eq_H_FM})] in the absence of the cavity. In the ground state, spins are aligned along the $z$ axis for $H_0=0$ and tilted away from the $z$ axis toward the $y$ axis for finite $H_0$.  
To determine the ground state, we regard $\hat{\bm{S}}_{\bm{r}}$ as an ${\bm{r}}$-independent classical vector and minimize the mean-field energy with respect to the vector direction~\cite{suppl_ref}. This calculation reveals that there is an Ising phase transition due to the spontaneous symmetry breaking of the $\mathcal{M}_y$ symmetry at $\chi=1$ where $\chi=|\gamma \mu H_{0}|/2KS$. While the broken symmetry phase with a two-fold ground-state degeneracy appears in the weak-field regime ($\chi <1$), the symmetric phase with a unique ground state is realized in the strong-field regime ($\chi >1$).

To study low-energy magnon excitations above the ground state in each regime, we describe the deviation from the mean-field ground state using magnon operators. The spin operator is rotated from the global coordinates to the local coordinates and transformed into the magnon operator using the conventional linear spin wave theory~\cite{holstein1940field}. The resulting magnon Hamiltonian is given as follows in the Nambu basis $\Psi_{\bm{k}}=(a_{\bm{k}},a^{\dagger}_{-\bm{k}})^{\text{T}}$ \cite{suppl_ref},
\begin{equation}
\label{magnonH}
    \mathcal{H}_{\text{s}} = \sum_{\bm{k}} \Psi^{\dagger}_{\bm{k}}(A_{\bm{k}}I_{2}+B \sigma_{x}) \Psi_{\bm{k}}, 
\end{equation}
where $a_{\bm{k}}$ is the magnon operator.
$I_{2}$ is the identity, and $\sigma_{x}$ is the Pauli matrix in the Nambu space. 
$A_{\bm{k}}$ and $B$ have different expressions depending on the regime:
\begin{equation}
    A_{\bm{k}} = JSn_{1}\zeta_{\bm{k}} + \frac{KS}{2}f^{(1)}_{\chi},\: B= \frac{KS}{2}f^{(2)}_{\chi},
\end{equation}
where $\{f^{(1)}_{\chi},f^{(2)}_{\chi}\}= \{2-\chi^{2},\chi^{2}\}$ in the weak-field regime, and $\{f^{(1)}_{\chi},f^{(2)}_{\chi}\}= \{2\chi-1,1\}$ in the strong field regime. Here, $n_{1}$ is the coordination number, $\zeta_{\bm{k}}= (1-\frac{1}{n_{1}}\sum_{\pm\bm{\delta}} e^{i\bm{k}\cdot\bm{\delta} })$ and $\bm{\delta}$ is the displacement vector to the nearest neighbors. Note that $\mathcal{H}_{\text{s}}$ contains squeezing terms, which create or annihilate two magnons simultaneously. Their magnitudes are proportional to $B$ and thus determined by $K$ and $H_0$. $\mathcal{H}_{\text{s}}$ can be diagonalized as $\mathcal{H}_{\text{s}}=\sum_{\bm{k}} \Omega_{\bm{k}} \tilde{a}^{\dagger}_{\bm{k}}\tilde{a}_{\bm{k}}$ with the energy $\Omega_{\bm{k}}=2\sqrt{A^{2}_{\bm{k}}-B^{2}}$ by using the Bogoliubov transformation: $\tilde{a}_{\bm{k}}= a_{\bm{k}}\cosh{r_{\bm{k}}} + a^{\dagger}_{-\bm{k}}\sinh{r_{\bm{k}}}$ where $\tilde{a}_{\bm{k}}$ is the squeezed magnon operator and $r_{\bm{k}}$ is the squeezing parameter with
\begin{equation}
    r_{\bm{k}} = \log \left( \frac{A_{\bm{k}}+B}{A_{\bm{k}}-B} \right)^{1/4}.
\end{equation}
The magnon energy $\Omega_{\bm{k}}$ is minimized at $\bm{k}=0$ and the magnon energy gap $\Omega_{\bm{k}=0}$ vanishes at $\chi=1$ [Fig.~\ref{FIG_1}(b)]. The squeezing parameter is maximized at $\bm{k}=0$ for any value of $\chi$ [Fig.~\ref{FIG_1}(c)].

We focus on the transition region. Near $\chi=1$, the single-mode squeezing parameter $r_{\bm{k}=0}$ diverges [Fig.~\ref{FIG_1}(d)], implying the diverging magnonic quantum fluctuations at the energy gap closing. The energy gap $\Omega_0$ and the magnon squeezing $\Delta q_0=\braket{q_0^2}-\braket{q_0}^2 =e^{2r_0}/4$ with $q_0\equiv(a_0+a_0^\dagger)/\sqrt{2}$ exhibit power law scalings to zero and infinity, respectively, with an exponent $1/2$, implying the universality class of fully connected systems~\cite{sachdev_2011,dicke1954coherence,lipkin1965validity}. Furthermore, large $r_{\bm{k}}$
alludes to large spin angular momentum $S^{(1)}_{\bm{k}}$~\cite{kamra2016super} of a squeezed single-magnon state $|1_{\bm{k}}\rangle = \tilde{a}^{\dagger}_{\bm{k}} |0\rangle$, since
$r_{\bm{k}}$ exponentially enhances $S^{(1)}_{\bm{k}}$ for large $r_{\bm{k}}$, 
\begin{equation}
    S^{(1)}_{\bm{k}} /\hbar = \langle 1_{\bm{k}} | a^{\dagger}_{\bm{k}}a_{\bm{k}} | 1_{\bm{k}} \rangle \simeq e^{2r_{\bm{k}}}/4.
\label{Eq_SAM_Mag}
\end{equation}
Finally, we remark that this mechanism of the magnon squeezing enhancement is robust to the higher-order contribution of the magnetocrystalline anisotropy~\cite{suppl_ref}.

{\it Enhanced cavity-magnon coupling.---} 
Having established a mechanism for the generation of squeezing in the easy-axis ferromagnet, we now analyze the cavity magnonics Hamiltonian $\mathcal{H}_{\text{tot}}$ [Eq.~(\ref{Eq_H_tot})]. We first consider a symmetric (normal) phase where the spin is aligned along the $y$ direction due to the strong static magnetic field and the cavity exhibits no photon condensation so that the $\mathcal{M}_{y}$ symmetry is conserved. In the normal phase, the effective Hamiltonian for the uniform magnonic mode $(a=a_{\bm{k}=0})$, which coherently couples to the cavity field, reads~\cite{suppl_ref}
\begin{equation}
\mathcal{H}_{\text{c}} = 
\Omega_0
\tilde{a}^{\dagger}\tilde{a} + \omega b^{\dagger}b -i \tilde{g}(\tilde{a}-\tilde{a}^{\dagger})(b+b^{\dagger}),
\label{Eq_Sq_Cavity}
\end{equation}
where $\Omega_0=KS/\sinh{2r_0}$ and 
\begin{equation}
\tilde{g}=ge^{r_0}.
\end{equation}
Remarkably, the effective cavity-magnon coupling strength $\tilde{g}$ is exponentially enhanced by the magnonic squeezing parameter $r_0$. 
This is one of our main results. The enhancement stems from the competition between the easy-axis anisotropy and the Zeeman interaction in $\mathcal{H}_{\text{FM}}$ [Eq.~(\ref{Eq_H_FM})]
and therefore can be tuned by the static magnetic field. We note that schemes to enhance the light-matter interaction have been proposed using squeezed cavity photons in time-dependent parametric driving setup~\cite{leroux2018enhancing,Zhu2020a,Qin2018a}. 
However, in the rotating frame where the enhanced coupling is realized, unwanted fast oscillating processes always exist 
and may limit the range of coupling strength that can be synthesized. Our scheme eliminates this issue by the magnon squeezing, an intrinsic property of the magnet, as a resource to enhance $\tilde{g}$ by tuning the {\it static} field. 

\begin{figure}
    \centering
    \includegraphics[width=\linewidth]{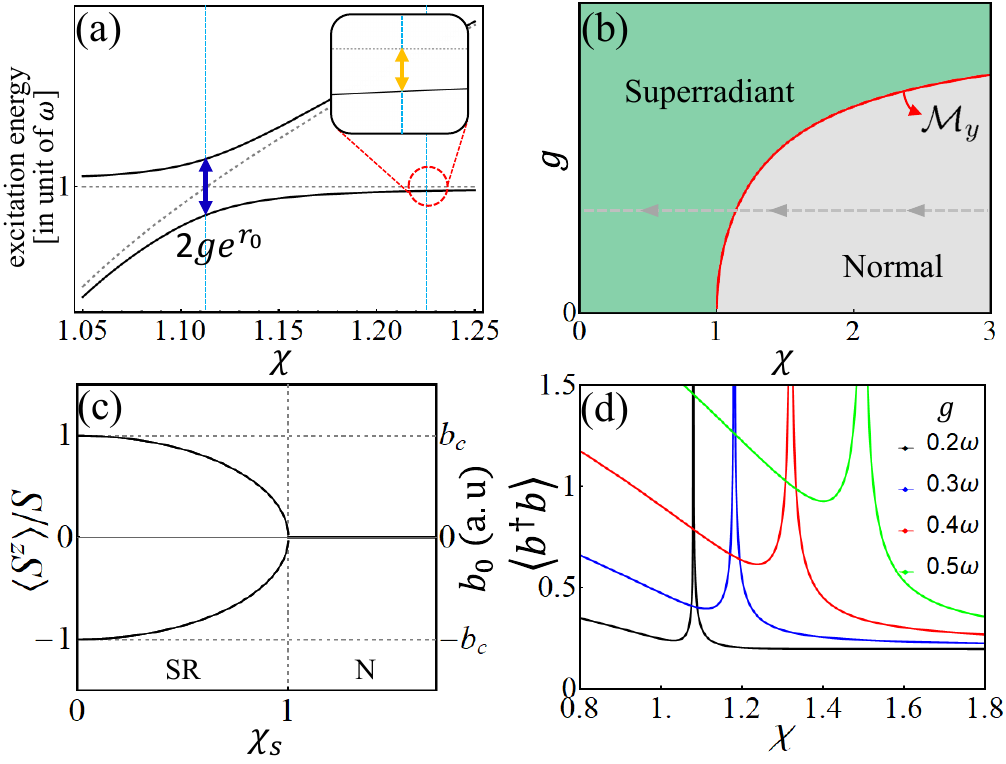}
    \caption{(a) Excitation energies of cavity magnonics Hamiltonian as a function of $\chi$ for a fixed coupling strength $g$. The blue and yellow arrows indicate the vacuum Rabi splitting and the dispersive cavity frequency shift, respectively. The dashed black line indicates the bare cavity and magnon frequencies without the interaction. (b) Phase diagram of the system along the parameter $\chi$ and the coupling strength $g$. (c) Order parameters $\langle S^{z}\rangle$ and $\langle b \rangle=b_0$ along the parameter $\chi$. $b_{c}$ denotes the maximum value of $\langle b \rangle$ for a fixed bare coupling strength $g$ and frequency $\omega$. The two graphs overlap each other. (d) The number of cavity photons $\langle b^{\dagger}b\rangle$ along the parameter $\chi$ for a few values of $g$.} 
    \label{FIG_2}
\end{figure}

Moreover, the enhanced $\tilde{g}$ allows a quantum optical measurement of the squeezing and the large spin angular momentum of magnons. To this end, one can tune $\tilde g$ to realize the strong coupling regime where $\tilde g$ is larger than the cavity (magnon) decay rate, $\kappa$ ($\Gamma$), but is still smaller than the cavity (magnon) frequency $\omega$ ($\Omega_{0})$; namely, $\kappa,\Gamma \ll \tilde g \ll \omega,\Omega_{0}$. On resonance, $\omega\sim\Omega_{0}$, an avoided crossing occurs giving rise to the vacuum Rabi splitting ($\Delta\omega$) that is determined by the magnon squeezing, i.e., $\Delta \omega \simeq 2ge^{r_{0}}$. In a dispersive limit, $g\ll|\omega-\Omega_{0}| $, the cavity frequency $\omega$ is shifted by $2g^{2}e^{2r_{0}}/\Delta \omega |_{g=0}$ where $\Delta \omega |_{g=0}$ is the detuning between the bare cavity and magnon frequencies. See Fig.~\ref{FIG_2}(a) for the vacuum Rabi splitting and the dispersive shift of the cavity magnonics system. Therefore, by measuring the frequency shift of the cavity field, which is a standard experimental tool in the cavity and circuit QED~\cite{blais2021circuit}, one can probe the degree of magnon squeezing. Note that the single excitation of magnon has successfully been probed using the same spectroscopic measurement~\cite{lachance2017resolving,lachance2020entanglement,xu2023quantum} and therefore our scheme is feasible with current technology.

Going beyond the strong coupling regime, one can also synthesize the ultrastrong coupling regime where the coupling strength becomes a dominant energy scale; therefore, exotic quantum optical phenomena that have been predicted with the ultrastrong light-matter interaction can be explored in our setup~\cite{Forndiaz2019a,FriskKockum2019a}. In particular, we find that there is a critical value of the squeezing parameter,
\begin{equation}
    r_{\text{max}}= \log \left(1 +\frac{KS \omega}{2g^{2}} \right)^{1/4},
\end{equation}
beyond which the normal phase Hamiltonian in Eq.~(\ref{Eq_Sq_Cavity}) becomes unstable. This indicates a possibility where the enhanced cavity-magnon coupling strength drives an SPT~\cite{Emary:2003da,Baumann:2010js}, which we examine in detail below.

{\it Magnonic superradiant phase transition.---} In order to examine the emergence of superradiant and magnetic order, we perform a mean-field analysis for $\mathcal{H}_\textrm{tot}$ in Eq.~(\ref{Eq_H_tot}) by considering $\hat{\bm{S}}_{\bm{r}}$ as an ${\bm{r}}$-independent classical vector and $b$ as a complex variable $\langle b\rangle = b_{0}$. Our analysis~\cite{suppl_ref} shows that the order parameters $\braket{S^z}$ for the magnetic order or the magnon condensation and $\braket{b}$ for the photon condensation indeed become non-zero when the dimensionless parameter $\chi_{s}=\chi(1+2g^2/KS\omega)^{-1}$ becomes smaller than the critical value 1~[Fig.~\ref{FIG_2}(b)].
In this superradiant phase, the $\mathcal{M}_{y}$ symmetry is spontaneously broken for the ground state, which leads to the two-fold degeneracy~[Fig.~\ref{FIG_2}(c)]. Note that the critical line, $\chi_{s}=1$, depends on both the bare coupling strength $g$ and $\chi$ [Fig.~\ref{FIG_2}(b)]. Therefore, both $\chi$ and $g$ modulations can, in principle, induce the SPT, which we call the magnonic SPT. 


Observing the $g$-induced SPT in equilibrium is challenging since $g$ arising from the magnetic dipole coupling between the cavity and the magnet~\cite{lachance2017resolving} is small and difficult to modulate in situ.
On the other hand, the $\chi$-induced magnonic SPT offers a promising and viable alternative to realize the SPT through simple static field adjustments. An additional advantage of the $\chi$-induced SPT is its independence from the influence of the so-called $A^2$ term, whose role for the emergence of the SPT remains controversial~\cite{Nataf:2010vl,viehmann2011superradiant,Vukics2014a,Bamba2014a,DiStefano2019a,Stokes2020a,andolina2020theory}, thanks to the relatively small bare coupling $g$. 
Furthermore, in a weak magnetic field, the presence of a superradiant phase is robust and stable since the fluctuation due to the light-matter interaction is not strong enough to break the ordering and the magnon squeezing of the ferromagnet. This strong stability is absent in the standard cavity QED systems. 
Therefore, our cavity-magnonic system offers an exciting opportunity to observe the SPT.

In the superradiant phase, the fluctuation around the order parameter is governed by the following Hamiltonian~\cite{suppl_ref},
\begin{equation}
    \begin{aligned}
        \mathcal{H}_{\text{sr}} &= \bar{\Omega} \bar{a}^{\dagger}\bar{a} + \omega \tilde{b}^{\dagger}\tilde{b} 
        - i \bar{g} (\tilde{b}+\tilde{b}^{\dagger})(\bar{a}-\bar{a}^{\dagger}), 
    \end{aligned}
    \label{Eq_Hsr}
\end{equation}
where
\begin{equation}
    \bar{\Omega} = \sqrt{ \Big(2KS+\frac{4g^{2}}{\omega}\chi_s^2 \Big)\Big(2KS\Big(1-\chi_s^2\Big)+\frac{4g^{2}}{\omega} \chi_s^2 \Big) },
\end{equation}
is the energy of the squeezed magnon $\bar{a}=a\cosh r_s+a^\dagger \sinh r_s$ with the squeezing parameter $r_{s}$, 
\begin{equation}
r_s=\log \left( \frac{KS+\frac{2g^2}{\omega}\chi_s^2}{KS(1-\chi_s^2)+\frac{2g^2}{\omega}\chi_s^2}\right)^{1/4},
\end{equation}
$\tilde{b}=b-\langle b \rangle$
is the redefined cavity photon, and $\bar{g}=g\chi_s e^{r_{s}}$ is the effective cavity-magnon coupling
strength. Note that $\bar{g}$ is also enhanced by the magnonic squeezing parameter $r_{s}$ in the superradiant phase.

We examine the critical properties near the magnonic SPT as $\chi$ (or $H_0$) is varied along the grey dashed line in Fig.~\ref{FIG_2}(b). Both $\mathcal{H}_\text{c}$ [Eq.~(\ref{Eq_Sq_Cavity})] and $\mathcal{H}_\text{sr}$ [Eq.~(\ref{Eq_Hsr})] have two eigenmodes. Among the two modes, the excitation energy of the magnon-like mode vanishes at the critical point, where both the magnon number $\langle a^\dagger a\rangle$ and the photon number $\langle b^\dagger b \rangle$ diverge due to the two-mode squeezing terms in $\mathcal{H}_\text{c}$ and $\mathcal{H}_\text{sr}$.
For instance, $\langle b^\dagger b \rangle$ diverges as $|\chi-\chi_\text{cr}|^{-1/2}$, where $\chi_\text{cr}=1+2g^2/KS\omega$ is the critical value of $\chi$~\cite{suppl_ref}. The variation of $\langle b^\dagger b \rangle$ with $\chi$ and $g$ is shown in Fig.~\ref{FIG_2}(d). The divergence of both $\langle a^\dagger a \rangle$ and $\langle b^\dagger b \rangle$ follows power-law behaviors whose exponents are the same as those of the SPT in the Dicke model, indicating that the magnonic SPT belongs to the universality class of fully connected systems~\cite{sachdev_2011,dicke1954coherence,lipkin1965validity,Emary:2003da}. The onset of the superradiant phase can be probed by the macroscopic occupation of the cavity photon due to the photon condensation, and the diverging fluctuation near the critical point can be probed by introducing a superconducting qubit dispersively coupled to the cavity field~\cite{Lolli2015a,schuster2007resolving,lachance2017resolving,lachance2020entanglement,blais2021circuit,xu2023quantum}.

{\it Estimations.---}
The experimental parameters are estimated using the yttrium iron garnet as a specific example since it is a popular choice for cavity magnonics experiments. In its bulk crystal, it exhibits a weak uniaxial anisotropy of approximately 80 Oe \cite{gieniusz1993cubic,lee2016ferromagnetic}. However, one can induce anisotropy by manipulating the sample geometry. For instance, its thin film can have an anisotropy field as $H_u \simeq 800$ Oe \cite{manuilov2009pulsed,wang2014strain,bhoi2018stress,bhoi2019observation}. This corresponds to an anisotropy coefficient $K \simeq 2.8$ GHz and the magnon frequency $\omega_{m}/2\pi \simeq 2.2$ GHz in the absence of the external magnetic field. Assuming a nanoscale sample with $1.8\cdot 10^8$ spins, the expected magnon-photon coupling strength is $g/2\pi \simeq 0.18$ GHz based on its unit coupling strength data \cite{tabuchi2014hybridizing,zhang2014strongly,bourhill2016ultrahigh}. Under these conditions, the critical field strength of the easy-axis magnet is estimated to be $H_0\simeq 810$ Oe and $H_0 \simeq 800$ Oe with or without the cavity, respectively, demonstrating the feasibility of observing the $\chi$-induced magnonic SPT. Moreover, we estimate that with a magnetic field of $H_0\simeq 795$ Oe, the cooperativity could increase about 20 times; therefore, it is feasible to synthesize both strong and ultrastrong coupling regimes of the cavity magnonics using our proposal.

{\it Conclusion.---} We have established the theory of magnon for the easy-axis ferromagnet and demonstrated that it endows the cavity-magnonics with a unique capability for enhancing the light-matter interaction using a \emph{static} magnetic field in stark contrast to the traditional cavity or circuit QED systems based on an ensemble of two-level systems where a time-dependent parametric driving is typically needed. The proposed cavity-magnonics system based on the easy-axis ferromagnet allows one 1) to probe the large magnon spin using quantum optical measurement toolboxes that are, in general, more robust against the effects of noises than the transport measurement and 2) to realize the magnonic SPT simply by tuning the static magnetic field strength without modulating the bare cavity-magnon coupling strength that is not large enough to drive the SPT by itself.

{\it Note added.---} We became aware of a recent work~\cite{bauer2023soft} where a similar prediction on the divergence of magnon squeezing in the easy-axis ferromagnet is reported.

\acknowledgements
J.M.L. and H.-W.L. acknowledge support from the Samsung Science and Technology Foundation (BA-1501-51). J.M.L. was supported by the POSCO Science Fellowship of the POSCO TJ Park Foundation. M.-J.H. was supported by the Startup Fund from Duke Kunshan University, and Innovation Program for Quantum Science and Technology 2021ZD0301602. J.M.L. thanks Min Ju Park and Ryan McKenzie for the fruitful discussions.

\bibliography{BibRef}

\end{document}


\title{Supplemental Material for ``Cavity magnonics with easy-axis ferromagnet: critically enhanced magnon squeezing and light-matter interaction"}

\author{Jongjun M. Lee}
\thanks{Electronic Address: michaelj.lee@postech.ac.kr}
\affiliation{Department of Physics, Pohang University of Science and Technology (POSTECH), Pohang 37673, Korea}

\author{Hyun-Woo Lee}
\thanks{Electronic Address: hwl@postech.ac.kr}
\affiliation{Department of Physics, Pohang University of Science and Technology (POSTECH), Pohang 37673, Korea}

\author{Myung-Joong Hwang}
\thanks{Electronic Address: myungjoong.hwang@duke.edu}
\affiliation{Division of Natural and Applied Sciences, Duke Kunshan University, Kunshan, Jiangsu 215300, China}
\affiliation{Zu Chongzhi Center for Mathematics and Computational Science, Duke Kunshan University, Kunshan, Jiangsu 215300, China}

\date{\today}
\maketitle
\tableofcontents
\appendix


\section{Detailed Calculation of the Critically Enhanced Magnon Squeezing}
In this section, we provide calculation details for the theory of magnon squeezing in the easy-axis ferromagnet.

\subsection{Derivation of the magnonic Hamiltonian}
We here derive the magnon Hamiltonian of the easy-axis ferromagnet under the static magnetic field in the main text. Starting from the spin Hamiltonian $\mathcal{H}_\textrm{FM}$ in Eq. (2) of the main text, the mean-field energy is given as follows with the classical uniform spin $\hat{\bm{S}}_{\bm{r}} = S(\sin\theta\cos\phi,\sin\theta\sin\phi,\cos\theta)$ and the magnetic field $\hat{\bm{H}}=H_{0}\hat{y}$.
\begin{equation}
    E_{\text{MF}}/N = JS^{2}n_{1} - KS^{2}\cos^{2}{\theta} -\gamma \mu H_{0}S \sin{\theta}\sin{\phi},
\end{equation}
where $N$ is the number of spins in the magnet. By considering the first-order derivatives in $\theta$ and $\phi$, we find the minimization conditions for the two regimes: (i) weak-field regime $(0<\chi<1)$ with $(\sin\theta,\phi)=(\chi,\pi/2)$ and (ii) strong-field regime $(1<\chi)$ with $(\theta,\phi)=(\pi/2,\pi/2)$ where $\chi=|\gamma \mu H_{0}|/2KS$. For each regime, the mean-field spin is given by 
\begin{equation}
    \begin{aligned}
        \bm{S} &= S( \sqrt{1-\chi^{2}}\hat{y}+\chi\hat{z}),\: (\text{weak-field regime}), \\
        \bm{S} &= S\hat{y},\: (\text{strong-field regime}).
    \end{aligned}
\end{equation}
To expand the spin Hamiltonian along the mean-field spin direction, we rotate the spin operator into the local axis of the mean-field spin direction. 
\begin{equation}
    \begin{pmatrix}
        \hat{S}^{x}_{\bm{r}} \\ \hat{S}^{y}_{\bm{r}} \\ \hat{S}^{z}_{\bm{r}}
    \end{pmatrix} = 
    \begin{pmatrix}
        1 & 0 & 0 \\
        0 & \cos\theta & \sin\theta \\
        0 & -\sin\theta & \cos\theta 
    \end{pmatrix}
    \begin{pmatrix}
        \hat{S}^{x'}_{\bm{r}} \\ \hat{S}^{y'}_{\bm{r}} \\ \hat{S}^{z'}_{\bm{r}}
    \end{pmatrix} ,
\end{equation}
where $\theta$ is defined with the minimization condition for each regime. In the local frame, we transform the spin operator into the magnon operator by using the Holstein-Primakoff transformation: $S^{x'}_{\bm{r}}= \sqrt{S/2}(a_{\bm{r}}+a^{\dagger}_{\bm{r}}),\: S^{y'}_{\bm{r}}= -i\sqrt{S/2}(a_{\bm{r}}-a^{\dagger}_{\bm{r}}),\: S^{z'} = S- a^{\dagger}_{\bm{r}}a_{\bm{r}}$, where $a_{\bm{r}}$ is the magnon operator at $\bm{r}$~\cite{holstein1940field}. First, in the strong field regime, the derivation is straightforward. By performing the Holstein-Primakoff transformation with $\theta= \pi/2$ and the Fourier transformation with
\begin{equation}
    a_{\bm{r}} = \frac{1}{\sqrt{N}}\sum_{\bm{k}} e^{i\bm{k}\cdot\bm{r}} a_{\bm{k}},
\end{equation}
one gets the following Hamiltonian,
\begin{equation}
    \mathcal{H}_{\text{s}} = \sum_{\bm{k}} \Psi^{\dagger}_{\bm{k}}(A_{\bm{k}}I_{2}+B \sigma_{x}) \Psi_{\bm{k}}, 
\end{equation}
where $A_{\bm{k}}=JSn_{1}\zeta_{\bm{k}}+KS(2\chi-1)/2$, $B=KS/2$, $\Psi_{\bm{k}}=(a_{\bm{k}},a^{\dagger}_{-\bm{k}})^{\text{T}}$, $I_{2}$ is the identity, and $\sigma_{x}$ is the Pauli matrix in the Nambu space. Here, $n_{1}$ is the coordination number and $\zeta_{\bm{k}}= (1-\frac{1}{n_{1}}\sum_{\pm\bm{\delta}} e^{i\bm{k}\cdot\bm{\delta} })$.

Second, in the weak field regime, the resulting Hamiltonian after the same transformations is a bit more involved, and therefore we provide a detailed derivation in the following. The exchange interaction simply reads
\begin{equation}
    -JS^{2}n_{1}N + 2JSn_{1}\sum_{\bm{k}} \Big(1-\frac{1}{n_{1}}\sum_{\pm\bm{\delta}}e^{i\bm{k}\cdot\bm{\delta}} \Big) a^{\dagger}_{\bm{k}}a_{\bm{k}}.
\end{equation}
The anisotropy term is after the Holstein-Primakoff transformation is given by, 
\begin{equation}
    \begin{aligned}
        &-K \sum_{\bm{r}} \Big( i\sin\theta \sqrt{\frac{S}{2}} (a_{\bm{r}}-a^{\dagger}_{\bm{r}})+ \cos\theta(S-a^{\dagger}_{\bm{r}}a_{\bm{r}})  \Big)^{2}
        -\gamma' H_{0} \sum_{\bm{r}} \Big( -i\cos\theta \sqrt{\frac{S}{2}} (a_{\bm{r}}-a^{\dagger}_{\bm{r}})+ \sin\theta(S-a^{\dagger}_{\bm{r}}a_{\bm{r}}) \Big) \\
        \simeq& -K \sum_{\bm{r}} \Big[ -\sin^{2}\theta \frac{S}{2} (a_{\bm{r}}-a^{\dagger}_{\bm{r}})^{2} + \cos^{2}\theta (S^{2}-2Sa^{\dagger}_{\bm{r}}a_{\bm{r}} )
        +2i\sin\theta \cos\theta S\sqrt{\frac{S}{2}} (a_{\bm{r}}-a^{\dagger}_{\bm{r}})  \Big] \\
        &-\gamma' H_{0} \sum_{\bm{r}} \Big[ -i\cos\theta \sqrt{\frac{S}{2}}(a_{\bm{r}}-a^{\dagger}_{\bm{r}}) + \sin\theta (S-a^{\dagger}_{\bm{r}}a_{\bm{r}}) \Big],
    \end{aligned}
\end{equation}
where higher-order terms of the magnon operators were neglected and $\gamma'=\gamma \mu$. If we substitute the minimization condition of $\theta$, then the terms proportional to $(a_{\bm{r}}-a^{\dagger}_{\bm{r}})$ vanish each other, i.e.
\begin{equation}
      \sum_{\bm{r}} \Big[ -2KS \sin\theta    
       +\gamma' H_{0}\Big]i\cos\theta \sqrt{\frac{S}{2}}(a_{\bm{r}}-a^{\dagger}_{\bm{r}})=0.
\end{equation}
Then, after the Fourier transformation, the Hamiltonian is given by,
\begin{equation}
\begin{aligned}
-KS^{2}N \cos^{2}\theta -\gamma' NS H_{0} \sin\theta -KS\sum_{\bm{k}} \Big[ -\frac{\sin^{2}\theta}{2} (a_{\bm{k}}-a^{\dagger}_{-\bm{k}})(a_{-\bm{k}}-a^{\dagger}_{\bm{k}})-2\cos^{2}\theta a^{\dagger}_{\bm{k}}a_{\bm{k}} \Big] +\gamma' H_{0} \sin\theta \sum_{\bm{k}}   a^{\dagger}_{\bm{k}}a_{\bm{k}}.
\end{aligned}
\end{equation}
By substituting the minimization condition for $\theta$ and neglecting the constant terms, one can derive the following magnon Hamiltonian of the weak-field regime.
\begin{equation}
    \mathcal{H}_{\text{s}} = \sum_{\bm{k}} \Psi^{\dagger}_{\bm{k}}(A_{\bm{k}}I_{2}+B \sigma_{x}) \Psi_{\bm{k}}, 
\end{equation}
where $A_{\bm{k}}=JSn_{1}\zeta_{\bm{k}}+KS(2-\chi^{2})/2$, $B=KS\chi^{2}/2$, $\Psi_{\bm{k}}=(a_{\bm{k}},a^{\dagger}_{-\bm{k}})^{\text{T}}$, $I_{2}$ is the identity, and $\sigma_{x}$ is the Pauli matrix. The results correspond to the Hamiltonian in the main text.

\subsection{Critical behavior of the physical values}
We diagonalize the Hamiltonian by using the Bogoliubov transformation $a_{\bm{k}}= \tilde{a}_{\bm{k}} \cosh{r_{0}}-\tilde{a}^{\dagger}_{-\bm{k}} \sinh{r_{0}}$ and investigate the critical behavior of the physical quantities near the critical point $\chi=1$. One can simply calculate the frequency and the squeezing parameter by using the formula in the main text. The frequency $\Omega_{\bm{k}}$ of the ground state at $\bm{k}=0$ is given by,
\begin{equation}
\begin{aligned}
    \Omega_{0} &= 2KS |1-\chi^{2}|^{1/2} \propto |1-\chi|^{1/2},\: \text{(weak-field regime)},\\
    \Omega_{0} &= 2KS |\chi(\chi-1)|^{1/2} \propto |1-\chi|^{1/2},\: \text{(strong-field regime)}.
\end{aligned}
\end{equation}
The squeezing parameter at $\bm{k}=0$ is given by,
\begin{equation}
\begin{aligned}
    e^{r_{0}} &= |1-\chi^{2}|^{-1/4} \propto |1-\chi|^{-1/4},\: \text{(weak-field regime)},\\
    e^{r_{0}} &= \chi^{1/4} |\chi-1|^{-1/4} \propto|1-\chi|^{-1/4},\: \text{(strong-field regime)}.
\end{aligned}
\end{equation}
The critical behavior corresponds to the numerical results in the main text.

\begin{figure}
    \centering
    \includegraphics[width=14cm]{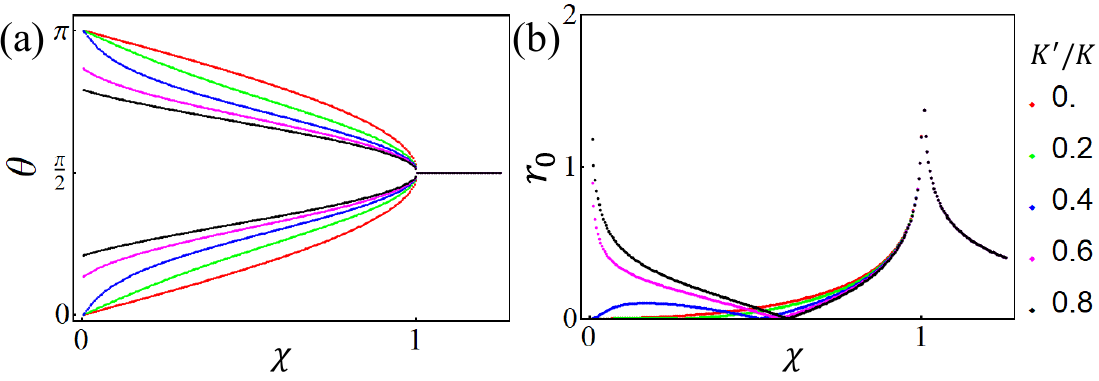}
    \caption{(a) The angle $\theta$ and (b) the single-mode squeezing parameter $r_0$ of the ground state condition as a function of the parameter $\chi$ for several ratios of $K'/K$.}
    \label{fig_supp_1}
\end{figure}

\subsection{Higher-order term of the magnetocrystalline anisotropy}
We consider the higher-order contribution of the magnetocrystalline anisotropy combined with the uniaxial quadratic-order contribution. It induces the Hamiltonian given by
\begin{equation}
    \mathcal{H}_{\text{h}} = -J \sum_{\langle \bm{r}, \bm{r'}\rangle} \hat{\bm{S}}_{\bm{r}} \cdot \hat{\bm{S}}_{\bm{r'}}
     - \sum_{\bm{r}}(K \hat{S}^{z2}_{\bm{r}} + \gamma' \hat{\bm{H}} \cdot \hat{\bm{S}}_{\bm{r}} ) + K' \sum_{\bm{r}} S^{z4}_{\bm{r}}, 
\end{equation}
where $K'>0$ is the strength of the quartic-order term~\cite{cullity2011introduction}. When the field is absent ($\chi=0$), the system has an easy axis along the $z$ axis if $K>2K'S^{2}$ and an easy cone along the $z$ axis if $K<2K'S^{2}$. To investigate the effect of the quartic order contribution in the presence of the field, we perform mean-field theory and find the ground state condition of $\theta$ as a function of the parameter $\chi$. The results are plotted in FIG.~\ref{fig_supp_1}(a), which shows that the phase transition still occurs at $\chi=1$ regardless of the quartic-order anisotropy. Although the quartic-order term adds quantitative corrections to the function $\theta(\chi)$, we see that the phase transition behavior remains qualitatively unchanged near the critical point. Furthermore, the regime with the easy cone ($K'/K=0.6,\:0.8$ in FIG.~\ref{fig_supp_1})
still exhibits the same transition point $\chi=1$, although its stable solution at $\chi=0$ is no longer along the $z$ axis. We additionally plot the squeezing parameter as a function of $\chi$ in FIG.~\ref{fig_supp_1}(b). One can see that the squeezing parameter still diverges at the same critical point regardless of the quartic term. The quartic term adds quantitative corrections to the function $r_0(\chi)$; for example, there is a non-zero value of $\chi$ at which the squeezing parameter becomes zero. Moreover, in the regime of the easy cone ($K'/K=0.6,\:0.8$ in FIG.~\ref{fig_supp_1}), one can see that the squeezing parameter diverges without the magnetic field. We leave the magnonic nature in the easy cone regime, which exhibits additional singular behaviors, as a future study. These findings demonstrate that the presence of the phase transition and the mechanism for the generation of critically enhanced squeezing is robust to the quartic-order term of the anisotropy.


\section{Detailed Derivation of the Cavity Magnonics Hamiltonian}
We consider the spin-cavity Hamiltonian in the main text:
\begin{equation}
    \mathcal{H}= -J \sum_{\langle \bm{r}, \bm{r'}\rangle} \hat{\bm{S}}_{\bm{r}} \cdot \hat{\bm{S}}_{\bm{r'}}
     - \sum_{\bm{r}}(K \hat{S}^{z2}_{\bm{r}} + \gamma' H_{0} \cdot \hat{S}^{y}_{\bm{r}} ) +\omega b^{\dagger}b - \frac{g}{\sqrt{N}}\sqrt{\frac{2}{S}} (b+b^{\dagger})\sum_{\bm{r}} \hat{S}^{z}_{\bm{r}},
\end{equation}
where the anisotropy and the cavity field are parallel to each other. Let us investigate the ground state of the Hamiltonian by deriving the cavity magnonics Hamiltonian with the general local coordinates parametrized by $\theta$,
\begin{equation}
    \begin{pmatrix}
        \hat{S}^{x}_{\bm{r}} \\ \hat{S}^{y}_{\bm{r}} \\ \hat{S}^{z}_{\bm{r}}
    \end{pmatrix} = 
    \begin{pmatrix}
        1 & 0 & 0 \\
        0 & \cos\theta & \sin\theta \\
        0 & -\sin\theta & \cos\theta 
    \end{pmatrix}
    \begin{pmatrix}
        \hat{S}^{x'}_{\bm{r}} \\ \hat{S}^{y'}_{\bm{r}} \\ \hat{S}^{z'}_{\bm{r}}
    \end{pmatrix},
\end{equation}
that minimizes the mean-field energy. In the rotated coordinate, we perform the Holstein-Primakoff transformation and the Fourier transformation. The exchange interaction is given by 
\begin{equation}
    \mathcal{H}_{J} = -JS^{2}Nn_{1} + 2JSn_{1}\sum_{\bm{k}} \Big(1-\frac{1}{n_{1}}\sum_{\pm\bm{\delta}}e^{i\bm{k}\cdot\bm{\delta}} \Big) a^{\dagger}_{\bm{k}}a_{\bm{k}}.
\end{equation}
Note that it is invariant under rotation since it is assumed to be isotropic. The anisotropy term is given by
\begin{equation}
    \begin{aligned}
        \mathcal{H}_{K} =& -KNS^{2} \cos^{2}\theta -2iKS \sin\theta\cos\theta \sqrt{\frac{NS}{2}} (a-a^{\dagger})+ \frac{KS}{2}\sin^{2}\theta \sum_{\bm{k}}(a_{\bm{k}}-a^{\dagger}_{-\bm{k}})(a_{-\bm{k}}-a^{\dagger}_{\bm{k}}) \\
        &+2KS \cos^{2}\theta \sum_{\bm{k}}a^{\dagger}_{\bm{k}}a_{\bm{k}}.
    \end{aligned}
\end{equation}
where we have defined the uniform magnonic mode, i.e., the Kittel mode, as $a\equiv a_{k=0}$.
The Zeeman interaction with the static field is given by
\begin{equation}
    \begin{aligned}
        \mathcal{H}_{Z} =& - \gamma' H_{0}NS\sin\theta +i\gamma' H_{0}\sqrt{\frac{NS}{2}}\cos\theta (a-a^{\dagger})+ \gamma' H_{0}\sin\theta \sum_{\bm{k}}a^{\dagger}_{\bm{k}}a_{\bm{k}} .
    \end{aligned}
\end{equation}
And the cavity-related terms are given by
\begin{equation}
    \mathcal{H}_{\text{cav}} = \omega b^{\dagger}b -g \sqrt{\frac{2}{NS}} (b+b^{\dagger}) \Big[ i\sin\theta \sqrt{\frac{NS}{2}} (a-a^{\dagger})+\cos\theta(NS-\sum_{\bm{k}} a^{\dagger}_{\bm{k}}a_{\bm{k}}) \Big].
\end{equation}
Since the cavity field only couples to the uniform magnonic mode, we keep the mode $a$
\begin{equation}
    \begin{aligned}
        \mathcal{H} =& -JS^{2}Nn_{1} -KNS^{2}\cos^{2}\theta -\gamma' H_{0}NS\sin\theta + \omega b^{\dagger}b -2iKS \sin\theta \cos\theta \sqrt{\frac{NS}{2}}(a-a^{\dagger}) \\
        &+i\gamma' H_{0}\sqrt{\frac{NS}{2}} \cos\theta (a-a^{\dagger}) -ig\sin\theta (b+b^{\dagger})(a-a^{\dagger}) + \frac{KS}{2} \sin^{2}\theta (a-a^{\dagger})^{2} + 2KS\cos^{2}\theta a^{\dagger}a \\
        &+ \gamma' H_{0}\sin\theta a^{\dagger}a,
    \end{aligned}
\end{equation}
where higher-order terms of the magnon operator were neglected. With the arbitrary magnon-photon coupling strength, both the magnon and the photon potentially become condensed as a coherent state. To find the mean-field solution, we assume the following conditions for the magnon and the cavity photon.
\begin{equation}
    \langle a \rangle = \langle a^{\dagger} \rangle =0 ,\: \langle b \rangle = \langle b^{\dagger}\rangle = b_0.
\end{equation}
Note that the magnon condensation is implicitly considered as a rotation of the mean-field spin. Then, we have the mean-field energy,
\begin{equation}
    \begin{aligned}
        E_{\text{MF}} =& -JS^{2}Nn_{1} -KS^{2}N \cos^{2}\theta -2g\sqrt{2NS} b_{0}\cos\theta - H_{0}NS \sin\theta +\omega b^{2}_{0}.
    \end{aligned}
\end{equation}
The minimization conditions are $b_{0}=(g\sqrt{2NS}/\omega)\cos\theta$ and
\begin{equation}
    \begin{gathered}
        \sin{\theta}= \chi_{s},\: b_{0}=\frac{g\sqrt{2NS}}{\omega}\sqrt{1-\chi^{2}_{s}} ,\: (\chi_{s}<1), \\
        \sin{\theta}= \frac{\pi}{2},\: b_{0}=0 ,\:  (\chi_{s}>1),
    \end{gathered}
\end{equation}
where $\chi_{s}=\chi /(1+2g^{2}/KS \omega)$. 
The conditions are critical at $\chi_{s}=1$. In the first regime, referred to as the superradiant phase, the photon is condensed. However, in the second regime, referred to as the normal phase, no such condensation occurs. The ground state is in the superradiant (normal) phase if $g>g_{c}$ ($g<g_{c}$) where $g_c = \sqrt{2KS\omega(\chi-1)}/2$. In the normal phase, the Hamiltonian is given by,
\begin{equation}
    \mathcal{H} = KS(2 \chi -1)a^{\dagger}a + \omega b^{\dagger}b + \frac{KS}{2} (a^{2}+a^{\dagger 2}) - ig (a-a^{\dagger})(b+b^{\dagger}),
\end{equation}
Using the Bogoliubov transformation $a= \cosh{r_{0}}\tilde{a} - \sinh{r_{0}}\tilde{a}^{\dagger}$, we eliminate the magnonic single-mode squeezing term,
\begin{equation}
    \mathcal{H}= \Omega_{0} \tilde{a}^{\dagger}\tilde{a}  + \omega b^{\dagger}b - i\tilde{g} (\tilde{a}-\tilde{a}^{\dagger})(b+b^{\dagger}),
\label{Eq_Ham_N}
\end{equation}
where $\tilde{g}=ge^{r_{0}}$, $r_{0}= -\frac{1}{4} \log (1-\chi^{-1})$, and $\Omega_0 = KS/\sinh{2r_{0}}$. On the other hand, in the superradiant phase, the Hamiltonian is given by,
\begin{equation}
    \begin{aligned}
        \mathcal{H} =& \Big[ KS(2-\chi^{2}_{s})+ \frac{4g^{2}}{\omega}\chi^{2}_{s} \Big]a^{\dagger}a + \omega \tilde{b}^{\dagger}\tilde{b} + \frac{KS}{2}\chi^{2}_{s}(a^{2}+a^{\dagger 2}) -ig\chi_{s} (a-a^{\dagger})(\tilde{b}+\tilde{b}^{\dagger}),
    \end{aligned}
\end{equation}
where $\tilde{b}=b-b_{0}$. Similarly, using the Bogoliubov transformation $a= \cosh{r_{s}}\bar{a} - \sinh{r_{s}}\bar{a}^{\dagger}$, we eliminate the magnonic single-mode squeezing term,
\begin{equation}
    \mathcal{H} = \bar{\Omega} \bar{a}^{\dagger}\bar{a}  + \omega \tilde{b}^{\dagger}\tilde{b} - i\bar{g} (\bar{a}-\bar{a}^{\dagger})(b+b^{\dagger}),
\end{equation}
where $\bar{g}=e^{r_{s}}g\chi_{s}$, the squeezing parameter is
\begin{equation}
    e^{-4r_{s}} = 1-\frac{2KS\chi^{2}_{s}}{2KS + \frac{4g^{2}}{\omega}\chi^{2}_{s}},
\end{equation}
and the frequency is 
\begin{equation}
    \bar{\Omega} = \sqrt{ \Big( 2KS +\frac{4g^{2}}{\omega}\chi^{2}_{s} \Big)\Big( 2KS(1-\chi^{2}_{s}) +\frac{4g^{2}}{\omega}\chi^{2}_{s} \Big) }.
\end{equation}
Both Hamiltonians correspond to the Hamiltonians in the main text, and their magnon-photon coupling strength is enhanced by the magnonic squeezing.

\section{Diagonalization of the Cavity Magnonics Hamiltonian}
Consider two interacting bosonic modes that are the generalized form of the cavity magnonics Hamiltonian.
\begin{equation}
    \mathcal{H} = \omega_{1} a^{\dagger}a + \omega_{2} b^{\dagger}b + g(a+a^{\dagger})(b+b^{\dagger}),
\end{equation}
for example, $a$ is the photon operator, and $b$ is the magnon operator. We first consider the simplest case with $\omega_{1}= \omega_{2}= \omega$. Using the unitary transformation,
\begin{equation}
\begin{pmatrix}
    a \\ b
\end{pmatrix}
= \frac{1}{\sqrt{2}}
\begin{pmatrix}
    1 & -1 \\
    1 & 1
\end{pmatrix}
\begin{pmatrix}
    \alpha \\ \beta
\end{pmatrix},
\end{equation}
that decouples the two modes, we obtain
\begin{equation}
    \mathcal{H} = \omega \alpha^{\dagger}\alpha + \frac{g}{2}(\alpha+\alpha^{\dagger})^{2} 
    + \omega \beta^{\dagger}\beta - \frac{g}{2} (\beta+\beta^{\dagger})^{2}.
\end{equation}
Two decoupled fields can be diagonalized by calculating two independent Bogoliubov transformations,
\begin{equation}
\begin{aligned}
    \alpha = \cosh r_{\alpha} \tilde{\alpha}- \sinh r_{\alpha} \tilde{\alpha}^{\dagger}, \;
    \beta = \cosh r_{\beta} \tilde{\beta}+ \sinh r_{\beta} \tilde{\beta}^{\dagger},
\end{aligned}
\end{equation}
where the squeezing parameters are given as
\begin{equation}
    e^{r_{\alpha}} = \Big(\frac{\omega}{\omega+2g} \Big)^{1/4} ,\: e^{r_{\beta}} = \Big(\frac{\omega}{\omega-2g} \Big)^{1/4} .
\end{equation}
The resulting Hamiltonian is
\begin{equation}
    \mathcal{H} = \sqrt{\omega(\omega+2g)}\tilde{\alpha}^{\dagger}\tilde{\alpha} +\sqrt{\omega(\omega-2g)}\tilde{\beta}^{\dagger}\tilde{\beta}. 
\end{equation}
Note that one of the frequencies becomes imaginary if $g>\omega/2$. This gives the critical value of the coupling strength $g_{c}= \omega/2$. Then, one can simply investigate the critical behavior; for example, the gap of the lower frequency closes as $\omega_{-} \propto \sqrt{g_{c}-g}$. The number of excitations for each mode can be calculated as the following,
\begin{equation}
    \begin{aligned}
    \langle a^{\dagger}a \rangle =\langle b^{\dagger}b \rangle= \frac{1}{4} (\sinh^{2}{r_{\alpha}}+\sinh^{2}{r_{\beta}}) \propto (g_c -g)^{-1/2}.
    \end{aligned}
\end{equation}
Second, we consider a more general case for two interacting bosonic modes with $\omega_{1}\neq \omega_{2}$. One can diagonalize the Hamiltonian by using the generalized Bogoliubov transformation~\cite{tolkunov2007quantum},\begin{equation}
\begin{aligned}
    a &= u_{00} \alpha + v_{00}\alpha^{\dagger} + u_{01}\beta + v_{01} \beta^{\dagger}, \\ 
    b &= u_{10} \alpha + v_{10}\alpha^{\dagger} + u_{11}\beta + v_{11}\beta^{\dagger},
\end{aligned}
\end{equation}
where
\begin{equation}
    \begin{aligned}
    u_{00} &= \frac{\omega_{1}+\epsilon_{-}}{\sqrt{4\omega_{1}\epsilon_{-}}} \Big[ 1+ \frac{4g^{2}\omega_{1}\omega_{2}}{(\omega^{2}_{2}-\epsilon^{2}_{-})^{2}} \Big]^{-\frac{1}{2}}, \\
    u_{01} &= \frac{\omega_{1}+\epsilon_{+}}{\sqrt{4\omega_{1}\epsilon_{+}}} \Big[ 1+ \frac{4g^{2}\omega_{1}\omega_{2}}{(\omega^{2}_{2}-\epsilon^{2}_{+})^{2}} \Big]^{-\frac{1}{2}}, \\
    u_{10} &= - \frac{1}{\omega_{2}-\epsilon_{-}} \frac{\omega_{1} g}{\sqrt{\omega_{1}\epsilon_{-}}} \Big[ 1+ \frac{4g^{2}\omega_{1}\omega_{2}}{(\omega^{2}_{2}-\epsilon^{2}_{-})^{2}} \Big]^{-\frac{1}{2}},\\
    u_{11} &= - \frac{1}{\omega_{2}-\epsilon_{+}} \frac{\omega_{1}g}{\sqrt{\omega_{1}\epsilon_{+}}} \Big[ 1+ \frac{4g^{2}\omega_{1}\omega_{2}}{(\omega^{2}_{2}-\epsilon^{2}_{+})^{2}} \Big]^{-\frac{1}{2}},
    \end{aligned} 
\end{equation}
and
\begin{equation}
\begin{aligned}
    v_{00}= \frac{\omega_{1}-\epsilon_{-}}{\omega_{1}+\epsilon_{-}}u_{00},&\: v_{01}= \frac{\omega_{1}-\epsilon_{+}}{\omega_{1}+\epsilon_{+}}u_{01} \\
    v_{10}= \frac{\omega_{2}-\epsilon_{-}}{\omega_{2}+\epsilon_{-}}u_{10},&\: v_{11}= \frac{\omega_{2}-\epsilon_{+}}{\omega_{2}+\epsilon_{+}}u_{11}.
\end{aligned} 
\end{equation}
The resulting Hamiltonian reads
\begin{equation}
    \mathcal{H} = \omega_{-}\alpha^{\dagger}\alpha+ \omega_{+} \beta^{\dagger}\beta,
\end{equation}
where 
\begin{equation}
    \epsilon_{\pm} = \sqrt{ \frac{\omega^{2}_{1}+\omega^{2}_{2} \pm \sqrt{(\omega^{2}_{1}-\omega^{2}_{2})^{2}+16\omega_{1}\omega_{2}g^{2}}}{2} }.
\end{equation}
Note that $\epsilon_-$ closes its gap at $g_c=\sqrt{\omega_{1}\omega_{2}}/2$. One can calculate the number of photons for the ground state,
\begin{equation}
    \langle a^{\dagger} a \rangle = |u_{00}|^{2}+|u_{01}|^{2}.
\end{equation}
Near the critical point $g=g_c$, the photon number diverges with the square root function as follows,
\begin{equation}
    \langle a^{\dagger}a \rangle \simeq \frac{\sqrt{2g_{c}}\omega_{2}}{8\sqrt{\omega^{2}_{1}+\omega^{2}_{2}}} (g_{c}-g)^{-\frac{1}{2}},
\end{equation}
where $g_{c}=\sqrt{\omega_{1}\omega_{2}}/2$. 
If one substitutes the cavity magnonics Hamiltonian of the normal phase in Eq.~(\ref{Eq_Ham_N}), one can get the following results of the photon number as an example.
\begin{equation}
    \langle b^{\dagger}b\rangle \simeq \frac{\sqrt{2g_c} \Omega_{0}}{8\sqrt{\Omega^{2}_{0}+\omega^{2}}} \frac{1}{\sqrt{g_c -g}} 
    \simeq \frac{\Omega_{0}}{\sqrt{\Omega^{2}_{0}+\omega^{2}}} \frac{g_{c}}{4\sqrt{KS\omega/2}} \frac{1}{\sqrt{\chi-\chi_{\text{cr}}}},
\end{equation}
where $g_c = \frac{\sqrt{\Omega_{0}\omega}e^{-r_{0}}}{2}$ and $\chi_{\text{cr}}=1+2g^{2}/KS\omega$. This is the result given in the main text.

\bibliography{BibRef}